\documentclass[a4paper,11pt]{article}
\usepackage{pos}

\usepackage[T1]{fontenc}
\usepackage{graphicx}
\usepackage{epstopdf}
\usepackage{bm}
\usepackage{epsfig}
\usepackage{graphics}
\usepackage{xspace}
\usepackage{hyperref}
\usepackage{slashed}
\usepackage{xcolor}
\usepackage{longtable}
\usepackage[normalem]{ulem}
\usepackage{subcaption}
\usepackage{multirow}
\usepackage{orcidlink}
\usepackage{float}
\usepackage{booktabs}

\title{A Precise $\alpha_s$ Determination from the R-improved QCD Static Energy}

\author*{Jose M. Mena-Valle}

\affiliation{Departamento de F\'isica Fundamental and IUFFyM, Universidad de Salamanca\\
Plaza de la Merced S/N, E-37008 Salamanca, Spain}

\emailAdd{jmmena@usal.es}

\abstract{The strong coupling $\alpha_s$ is extracted with high precision through fits to lattice-QCD data for the static energy. Our theoretical framework is based on R-improving the three-loop fixed-order prediction for the static energy: we remove the $u=1/2$ renormalon and resum the associated large infrared logarithms. Combined with radius-dependent renormalization scales (the so-called profile functions), this procedure extends the range of validity of perturbation theory to distances as large as $\sim 0.5\,$fm. In addition, we resum large ultrasoft logarithms to N$^3$LL accuracy using renormalization-group evolution. Since the standard four-loop R-evolution treats N$^4$LL and higher-order contributions asymmetrically, we also incorporate this potential source of bias in our analysis. Our estimate of the perturbative uncertainty is obtained through a random scan over the parameters controlling the profile functions and the implementation of R-evolution. We analyze how the extracted value of $\alpha_s$ depends on the shortest and longest distances included in the fit, on the details of the R-evolution procedure, on the fitting strategy itself, and on the accuracy of ultrasoft resummation. From our final analysis, and after evolution to the $Z$ pole, we obtain $\alpha^{(n_f=5)}_s(m_Z)=0.1170\pm 0.0009$, a result fully compatible with the world average and with a comparable uncertainty.}

\FullConference{The European Physical Society Conference on High Energy Physics (EPS-HEP2025)\\
7-11 July 2025\\
Marseille, France\\}

\begin{document}
\maketitle

\section{Introduction}
The strong coupling constant $\alpha_s$ is the key parameter of Quantum Chromodynamics (QCD), governing the strength of the strong interaction and determining the energy scale where it becomes confining. Precise knowledge of $\alpha_s$ is essential for accurate predictions at lepton and hadron colliders and in other contexts such as Higgs decays, B-physics, and quarkonia. Although much progress has been made in reducing its uncertainty, different methods often yield inconsistent results, see Refs.~\cite{Salam:2017qdl,Pich:2018lmu,dEnterria:2022hzv,Gehrmann:2010uax,Abbate:2010xh,Abbate:2012jh,Gehrmann:2012sc,Hoang:2015hka,Bell:2023dqs,Benitez:2024nav,Benitez:2025vsp}. Among these, lattice QCD has become a leading approach, providing highly precise determinations \cite{FlavourLatticeAveragingGroupFLAG:2024oxs} that now dominate the PDG world average \cite{ParticleDataGroup:2024cfk}. One promising strategy that extracts $\alpha_s$ from lattice QCD simulations uses the QCD static energy, which is easy to compute on the lattice. It is the potential between a static heavy quark and antiquark at a distance $r$, dominated at short distances by UV dynamics and known perturbatively to $\mathcal{O}(\alpha_s^4)$. A lot of progress has been made since the first work of Ref.\cite{Bazavov:2012ka}, which used the data of \cite{Bazavov:2011nk}, by improving renormalon subtractions that lead to more precise determinations, as done in the HotQCD \cite{Bazavov:2014soa} (with data from \cite{Bazavov:2014pvz}) and TUMQCD \cite{Bazavov:2019qoo} analyses. In Ref.\cite{Mateu:2018zym}, the validity of the static potential was improved by using the MSR scheme \cite{Hoang:2008yj,Hoang:2017suc}, performing R-evolution, and using parametrized renormalization scales known as profile functions, being able to agree with HotQCD data \cite{Bazavov:2014soa} up to $1$\,fm. In this work, we perform fits to lattice data using our R-improved theoretical prediction of the static potential to extract the strong coupling and perform random scans varying the parameters that define our profile functions, designed to ensure that perturbation theory behaves well at large $r$ values. We also analyze how the dataset choice influences the result. Our $\chi^2$ function depends on the constants up to which the static energy for each lattice simulation is defined, so we have designed an optimized fitting algorithm that marginalizes the $\chi^2$ function with respect to them, making our procedure faster. We consider the cases where each ensemble has a different offset and where there is one common offset for all of them, and study how the results depend on either hypothesis. All derivations, technical details, and extended discussions
omitted here are presented in the long version of this work available on arXiv \cite{menavalle2025pot}.

\section{Theoretical Ingredients}
For our theoretical analysis we need the perturbative expression for the static potential. On top of that, we peform resummation of large ultrasoft logarithms, subtract the leading renormalon and carry out the associated R-evolution. All theoretical ingredients have been implemented in standalone \texttt{Python} and \texttt{C++} codes, which agree to about 15 decimal places. The \texttt{Python} version generates the interpolation
tables used in the fits, while the \texttt{C++} code directly minimizes the $\chi^2$ function. To RG evolve the MSR mass and $\alpha_s$, the \texttt{Python} code uses \texttt{REvolver}~\cite{Hoang:2021fhn}, while in \texttt{C++} we use a modified version of this package that depends on the perturbative result of $\Lambda_{\rm QCD}$. With either of them, we obtain the strong coupling with three active flavors at any scale from the reference value $\alpha_s^{(n_\ell=3)}(m_\tau)$
evolving with the five-loop $\beta$-function. To compute $\alpha_s^{(n_\ell=5)}(m_Z)$ we implement the charm and bottom threshold matching with four-loop accuracy.
A \texttt{SWIG}-based wrapper allows running the \texttt{C++} code within a \texttt{Python} environment, using \texttt{numpy}, \texttt{scipy}, and \texttt{matplotlib} for analyses and plotting. In our code and also the rest of this article, energies and distances are expressed in GeV and fm, using $\hbar c = 0.1973269804~{\rm GeV}\times{\rm fm}$ for the corresponding conversions.

The static energy $E_{\rm s}(r,\mu)$ is defined as the ground-state energy of a quark-antiquark pair at a distance $r$. It is scheme- and scale-independent, and can be obtain in potential NRQCD by adding the soft static potential $V_s(r,\mu)$, known up to three loops in fixed-order perturbation theory (N$^3$LO), to the ultrasoft contribution $\delta_{\rm us}(r,\mu)$ associated with the dynamical propagation of ultrasoft gluons that appear at $\mathcal{O}(\alpha_s^4)$. This contribution depends on the ultrasoft scale $\mu_{\text{us}}$ through the following logarithm: $L_{\rm us}=\log(C_A\alpha_s/(\mu_{\rm us}r))$. The static potential also has an ultrasoft contribution $V_{s}^{\rm us}(r,\mu)$ at the same order and its dependence with $\mu_{\rm us}$ is $L_s=\log(\mu_{\text{us}}re^{\gamma_E})$. Since there is no right choice for $\mu_{\text{us}}$ that minimizes both logarithms at the same time, one needs to perform resummation using the pNRQCD renomarlization group equations. We achieved N$^3$LL accuracy, that is, we are able to resumme all terms of the form $\alpha_s^{m}(\alpha_s L_{\rm us})^n$ for any $n\geq 0$ with $m=4$.

Subtracting the $\mathcal{O}(\Lambda_{\text{QCD}})$ renormalon ambiguity of the static potential in the $\overline{\mathrm{MS}}$ scheme is mandatory to achieve good convergence. This ambiguity is $r$-independent and equal to $-2$ times that of the series relating the pole and $\overline{\mathrm{MS}}$ masses for heavy quarks. In order to subtract the renormalon, we need to express the pole mass in terms of a short distance scheme, and our preferred choice is the MSR mass \cite{Hoang:2017suc}. The substraction series $\delta^{\mathrm{MSR}}$ must depend on a adjustable scale $R$, even though the ambiguity must not depend on it. Along with the other scales, we vary $R$ to minimize logarithms in $\delta^{\mathrm{MSR}}$, but this should not change the reference $R_0$ at which the potential is defined. This is achieved by solving the corresponding renormalization-group equation, known as R-evolution, which sums up the associated large logs. This process is called R-improvement \cite{Hoang:2009yr}, and more details of its implementation in the static potential can be found in Ref.~\cite{Mateu:2018zym}. The R-evolution kernel $\Delta^{\mathrm{MSR}}$ depends on the anomalous dimension of the MSR mass, which is not known up to infinite order. Hence, to account for the error of truncating at a finite order, we include a dimensionless parameter $\lambda$ in our analyses that will also be varied. The final expression for the static energy used in our code is the following:
\begin{align}\label{eq:Esum}
E_{\rm s}(r,\mu_s,\mu_{\rm us}) =\,& V^{\rm soft}_{\rm s}(r,\mu_s) +\frac{C_{\!A}^3C_{\!F}}{12}\,\frac{[\alpha_s(\mu_s)]^3}{r}
\biggl\{\frac{2}{\beta_0}\biggl(B-\frac{\beta_1}{4\beta_0}\biggr) \biggl[\frac{\alpha_s(\mu_{\rm us})}{\pi}-\frac{\alpha_s(\mu_s)}{\pi}\biggr]\\
&+\!\frac{2}{\beta_0}\biggl(1+3\frac{\alpha_s(\mu_s)}{4\pi} \Bigl[a_{10}+2\beta_0\log(r\mu_se^{\gamma_E})\Bigr]\biggr)\!
\log\biggl[\frac{\alpha_s(\mu_{\rm us})}{\alpha_s(\mu_s)}\biggr] -\frac{\alpha_s(\mu_s)}{\pi}L_r\nonumber \\
&-\!\frac{\alpha_s(\mu_{\rm us})}{\pi}\biggl[\log\biggl(\frac{C_{\!A}\alpha_s(\mu_s)}{r\mu_{\rm us}}\biggl)-\frac{5}{6}\biggr]
\biggr\}
+ 2 \delta^{\mathrm{MSR}}(R, \mu_s) + 2 \Delta^{\mathrm{MSR}}\left(R, R_0\right),\nonumber\\
V_{s}^{\rm soft}(r,\mu) =&-\! C_{\!F}\frac{\alpha_s(\mu)}{r}\sum_{i=0} \biggl[\frac{\alpha_s(\mu)}{4\pi}\biggr]^{i} \sum_{j=0}^{i} a_{ij} L_r^j\,, \quad a_{ij} = \frac{2}{j}\sum_{k=j}^{i}k\,a_{k-1,j-1}\beta_{i-k}\,.
\nonumber
\end{align}
In practice, it is only necessary to know the $a_{i0}$ coefficients ~\cite{Lee:2016cgz} since the rest, $a_{ij}$, can be computed through the recursive relation of the previous equation, first obtained in Ref.~\cite{Mateu:2017hlz}.

Finally, we need to choose a functional form for our renomarlization scales to keep small the logarithms and $\alpha_s(\mu)$ factors. The usual choice is the canonical profile used in \cite{Bazavov:2019qoo}, that describe very well the short distance domain but crosses the Landau pole for $r\gtrsim 1\,{\rm GeV}^{-1}\simeq 0.2\,$fm, spoiling good perturbative behavior. To solve this issue we use {\it profile functions}, as known in the context of event shapes ~\cite{Abbate:2010xh,Hoang:2014wka}. These are smooth $r$-dependent functions
that behave canonically ($\mu\sim 1/r$) at small $r$ and freeze at a certain value above the Ladau pole for larger values of the distance. We choose the following parametrization:
\begin{align}
\mu_s(r,\xi,\mu_\infty,\Delta,b) ={}& \sqrt{\biggl(\frac{\xi}{r}\biggr)^{\!\!2}+\frac{b}{r}+(\mu_\infty-\Delta)^2}-\Delta\,,
\\
R(r,\beta,R_\infty,\Delta,b) ={}& \mu_s(r,\beta,R_\infty,\Delta,b)\,,\nonumber \\
\mu_{\rm us}(r,\xi,\kappa,\mu_\infty,\Delta,b) ={}& C_A\Bigl\{\mu_s(r,\kappa\xi,\mu_\infty,\Delta,b)\alpha_s\bigl[\mu_s(r,\kappa\xi,\mu_\infty,\Delta,b)\bigr]-\mu_\infty\alpha_s(\mu_\infty)\Bigr\} + \mu_\infty\,.\nonumber
\end{align}
The different parameters defining the profile functions will be varied through a random scan to estimate perturbative uncertainties. Each one of them will take a random value inside variation ranges that are chosen so that no outliers occur and to ensure that $\mu_s>\mu_{\rm us}$ for small distances.

To perform our fits we use lattice QCD data from the HotQCD Collaboration published in Refs.~\cite{Bazavov:2014pvz,Bazavov:2016uvm,Bazavov:2017dsy} that use a (2+1)-flavor simulation. There are nine lattice sets, each one with a different lattice spacing and origin of the static energy that we call offset, totaling 2512 data points from $0$ to $1$\,fm. The value of $r_1$ used to convert from lattice to physical units is the weighted average of the results from Refs.~\cite{Larsen:2025wvg,Bazavov:2014pvz}, and we assign as uncertainty the regular average of the respective errors: $r_1=0.3093(20)\,$fm.

Our $\chi^2$ function, that includes our theoretical prediction and lattice uncertainties, depends on the various offsets. It is useful to minimize the $\chi^2$ function analytically with respect to them, obtaining a marginalized $\chi^2$ that only depends on $\alpha_s$. We carry out fits in two scenarios: employing a common offset for all datasets, or assuming an independent offset for each individual set. We produce a sample of 500 profiles with random parameters ---\,including $\lambda$\,--- and perform the minimization of the $\chi^2$ function for all of them. We end up with $500$ best fit values $\alpha_s^{\rm BF}$, all of them with an associated fit uncertainty, defined by \mbox{$\tilde\chi^2(\alpha_s^{\rm BF}\pm\Delta_{\rm fit}\alpha_s)=\chi^2_{\rm min}+1$}, which is the quadrature of the lattice and offset errors. We inflate the lattice error multiplying it by $\sqrt{\chi^2_{\rm min}/{\rm d.o.f.}}$ ( ${\rm d.o.f.}\equiv$ degrees of freedom). Our final central value (lattice uncertainty) is the average of the $500$ best-fit values (lattice uncertainties), and the perturbative uncertainty is defined by $\Delta_{\rm pert}\alpha_s=(\alpha_{\rm max} - \alpha_{\rm min})/2$.

\section{Fits for the Strong Coupling}

\begin{figure}[t!]
\centering
\begin{subfigure}{0.48\textwidth}
\centering
\includegraphics[width=\textwidth]{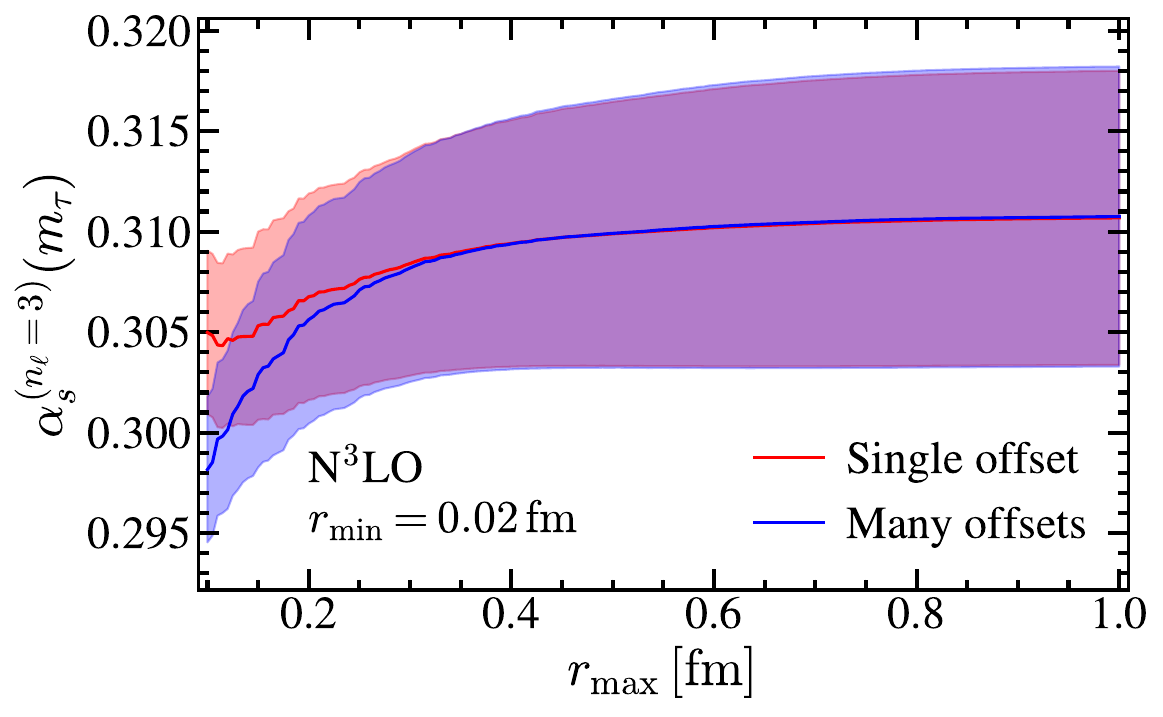}
\caption{}\label{fig:OffsetRange}
\end{subfigure}
\hfill
\begin{subfigure}{0.478\textwidth}
\centering
\includegraphics[width=\textwidth]{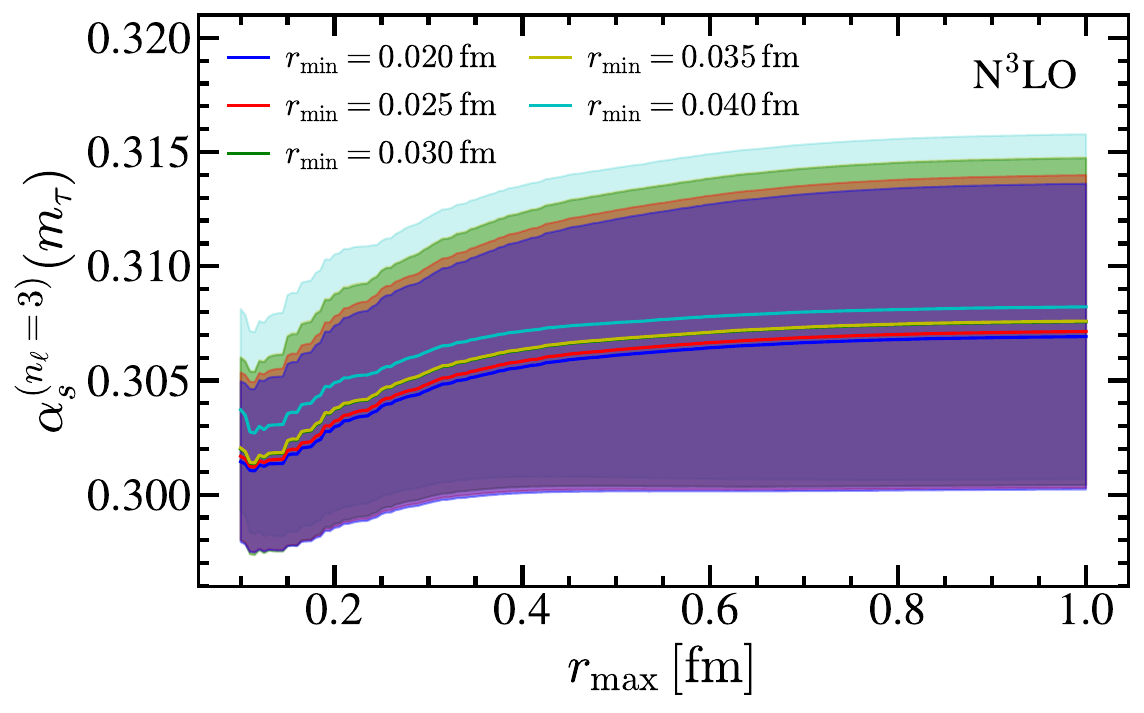}
\caption{}\label{fig:RangePlot}
\end{subfigure}

\caption{\label{fig:FinalOrders}
(a) Dependence of the strong coupling $\alpha_s$ on the maximal distance $r_{\rm max}$ included in the dataset for N$^3$LO fits, with $r_{\rm min}=0.02\,\mathrm{fm}$. Both the central value and the total uncertainty, which includes theoretical and lattice errors, are shown. Results with a single energy offset are shown in red, while multiple-offset fits are shown in blue. (b) N$^3$LO fit results for $\alpha_s$. Uncertainties include perturbative and lattice errors added in quadrature. The dependence of single-offset results on $r_{\rm max}$ is displayed for different $r_{\rm min}$ values: $0.02$, $0.025$, $0.03$, $0.035$, and $0.04\,\mathrm{fm}$, shown in blue, red, green, yellow, and cyan, respectively.}
\end{figure}

Before performing our final fit, we have studied different aspects of the dataset by varying the distance range $r_{\rm min}\leq r\leq r_{\rm max}$. First, we focus on the difference between considering the offsets of each lattice set as independent parameters in our fit versus a single offset for all the datasets. As seen in Fig.~\ref{fig:OffsetRange}, the two approaches show very good agreement for $r_{\max}> 0.3\,$fm whereas noticeable differences arise at shorter distances, where both the value and the uncertainty of $\alpha_s$ significantly reduce. We attribute the poor behavior of multi-offset fits to an overparameterization of the $\chi^2$ function, which can develop a runaway direction and push the extracted strong coupling to unnaturally small values.

Using the single offset approach, we change the value of $r_{\rm min}$ from $0.02$ to $0.04$\,fm. The results, shown in Fig.~\ref{fig:RangePlot}, indicate that the extracted value of $\alpha_s$ is sensitive to $r_{\rm min}$ and $r_{\rm max}$. In particular, larger values of either both of them lead to higher central values and larger uncertainties, reaching a plateau for $r_{\rm max} \gtrsim 0.4$\,fm. To account for the variation induced by these choices, we scan over \mbox{$r_{\rm min}\in[0.02,0.04]\,$fm} and $r_{\rm max}\in[0.35,0.45]\,$fm, and include a ``dataset'' uncertainty defined as half the difference between the largest and smallest central value obtained in this window. With all these requirements, our set consists in 105 elements. Taking all these considerations into account, our result at N$^3$LO order is
\begin{align}\label{eq:final3}
\!\alpha_s^{(n_\ell=3)}(m_\tau) ={}& 0.3093\pm 0.0061_{\rm th}\pm0.00001_{\rm lat}\pm 0.0011_{\rm set}\pm 0.0011_{r_1} \\
={} &0.3093\pm 0.0063_{\rm tot}\,,\nonumber
\end{align}
where we have also included the uncertainty on $r_1$. Finally, we need to evolve our result to the $Z$-pole. To do that we must cross the charm and bottom thresholds in order to match the strong coupling when going from three to four, and from four to five active flavors. This evolution is performed using the five-loop beta function and the four-loop matching conditions implemented in \texttt{REvolver}. The uncertainty associated with this procedure receives two contributions: a) the uncertainties in the charm and bottom masses, and b) variations of the matching scales. For the central values and errors of the charm and bottom quarks we take the PDG world averages ~\cite{ParticleDataGroup:2024cfk}. Care must be taking in choosing the matching scales: for the bottom one can use the standard factors of $2$ and $1/2$, but this cannot be implemented for the charm given its small mass. Instead, we use the standard factors of two on twice the charm mass, implying \mbox{$\mu_c=2^{1\pm1}\overline m_c$} and $\mu_b=2^{\pm1}\overline m_b$, where the notation $\overline m_q\equiv\overline m_q(\overline m_q)$ is in use. Adding all errors in quadrature we obtain
\begin{align}\label{eq:amZ}
\alpha_s^{(n_\ell=5)}(m_Z) ={} &0.1170\pm 0.0008_{\rm th}\pm 0.0001_{\rm set}\pm0.0001_{r_1}\pm 0.0003_{\mu_c}\pm0.0002_{\mu_b} \\
={} &0.1170\pm 0.0009_{\rm total}\,.\nonumber
\end{align}
\begin{figure}[t!]
\centering
\begin{subfigure}{0.49\textwidth}
\centering
\includegraphics[width=\textwidth]{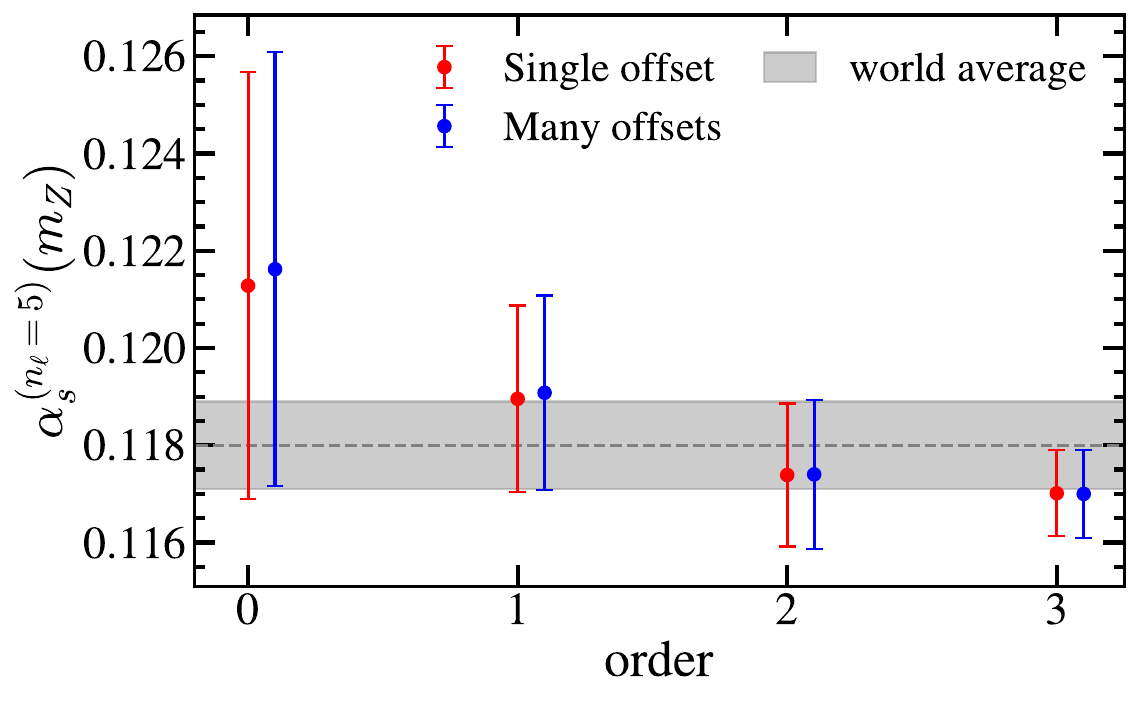}
\caption{}\label{fig:FinalAmZOrders}
\end{subfigure}
\hfill
\begin{subfigure}{0.47\textwidth}
\centering
\includegraphics[width=\textwidth]{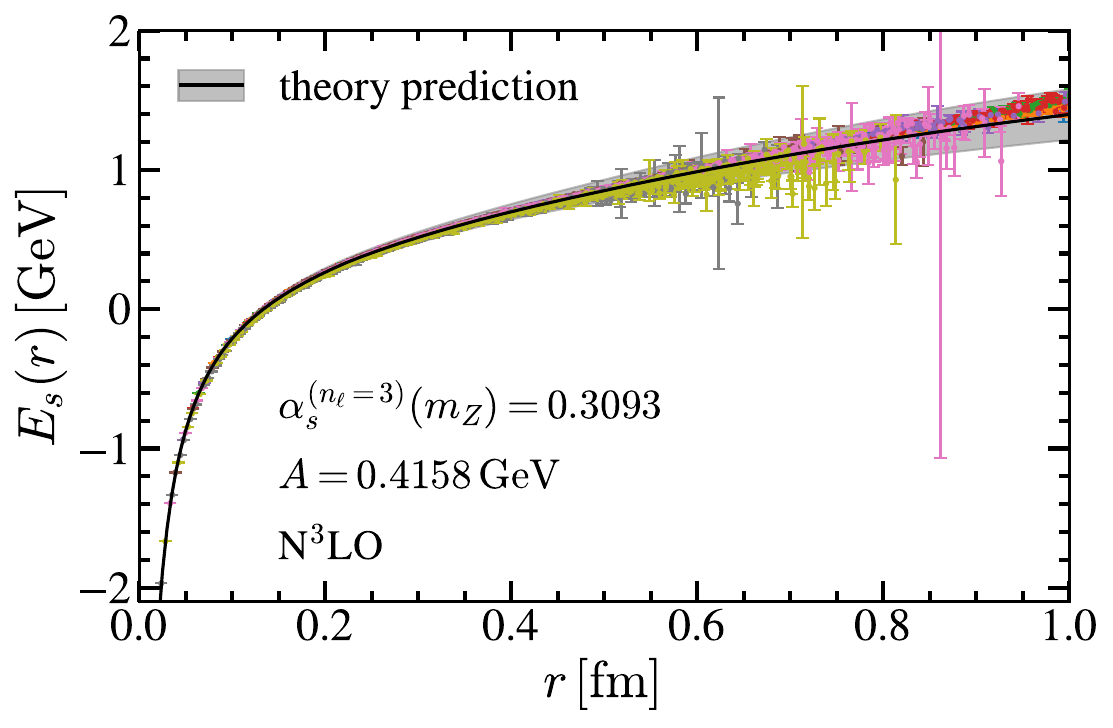}
\caption{}\label{fig:ThvsExp}
\end{subfigure}
\caption{\label{fig:FinalOrders2}
(a) Comparison of lattice data (colored dots with error bars) with our best theoretical prediction, using the central values for the strong coupling and energy offset from our final result in Eq.~\eqref{eq:final3}. Different colors indicate different lattice ensembles. The error bars show only lattice statistical uncertainties, while the gray bands represent the theoretical perturbative uncertainties. (b) Comparison of lattice data (colored dots with error bars) with our best theoretical prediction, using the central values of the strong coupling and energy offset from our final result in Eq.~\eqref{eq:final3}. Different colors indicate different lattice ensembles. The error bars show only lattice statistical uncertainties, while the gray bands represent the theoretical perturbative uncertainties}
\end{figure}
In Table \ref{tab:loops} we compare our results for different orders (only with the single offset approach) and in Fig.~\ref{fig:FinalAmZOrders} we compare them with the world average $0.1180(9)$. Nice order by order agreement and convergence are, as expected, displayed. The results for the multi-offset fits are also shown, finding that for higher orders the agreement between both methods is better. Finally, we can use our final value of $\alpha_s$ to compare the theoretical prediction for the static energy with the lattice data, applying the common offset obtained in our fit. The results shown in Fig.~\ref{fig:ThvsExp} depict very good agreement.

\section{Comparison with Previous Analyses}
It is useful to compare our final determination with recent extractions that also use the static energy. To be able to compare the different results we translate them to a common value of $r_1$ and ensure a consistent RG evolution from three flavors to the $Z$-pole.

Ref.~\cite{Ayala:2020odx} employs the interval $0.070\,\mathrm{fm} \leq r \leq 0.098\,\mathrm{fm}$, using the $\beta=8.4$ data from \cite{Bazavov:2019qoo} (8 points), and obtains $\alpha_s^{(n_\ell=5)}(m_Z)=0.1181(9)$, with $\alpha_s^{(n_\ell=3)}(m_\tau)=0.3151(65)$. Applying our theoretical framework to the same dataset yields $\alpha_s^{(n_\ell=3)}(m_\tau)=0.3123(80)$, indicating that our uncertainty estimate is more conservative. After converting to the updated $r_1$, their result becomes $\alpha_s^{(n_\ell=3)}(m_\tau)=0.3149(65)$. Evolving to five flavors with our prescription, setting the charm matching scale to twice their choice, we obtain $\alpha_s^{(n_\ell=5)}(m_Z)=0.1177(9)$, compatible with our result at the $0.55\sigma$ level.

The latest TUMQCD determination~\cite{Bazavov:2019qoo} uses points with $a\leq r \leq 0.073\,\mathrm{fm}$ (19 data points) and obtains $\alpha_s^{(n_\ell=5)}(m_Z)=0.1166(8)$ using $r_1$ from \cite{Bazavov:2014pvz}, they also quote $\alpha_s^{(n_\ell=3)}(m_\tau)=0.3039(62)$. Using our setup on the same dataset gives $\alpha_s^{(n_\ell=3)}(m_\tau)=0.3114(44)$. Using the prescription in Ref.~\cite{Ayala:2020odx} and evolving the TUMQCD 3-flavor strong coupling to the $Z$-pole the result for
$\alpha_s^{(n_\ell=5)}(m_Z)$ quoted in Ref.~\cite{Bazavov:2019qoo} is reproduced. Converting their value to the updated $r_1$ gives $\alpha_s^{(n_\ell=3)}(m_\tau)=0.3033$. Evolving to the $Z$-pole with our charm-threshold prescription we find $\alpha_s^{(n_\ell=5)}(m_Z)=0.1162(8)$, consistent with Eq.~\eqref{eq:amZ} within $0.63$-$\sigma$.

\begin{table}[t!]
\centering
\begin{tabular}{ c cccc } \toprule & LO & NLO & N$^2$LO & N$^3$LO \\\midrule
$\alpha_s^{(n_\ell=3)}(m_\tau)$ & $0.3458 \pm 0.0405$ & $0.3251 \pm 0.0158$ & $0.3123 \pm 0.0113$ & $0.3093 \pm 0.0063$\\[0.1cm]
$\alpha_s^{(n_\ell=5)}(m_Z)$ & $0.1213 \pm 0.0044$ & $0.1190 \pm 0.0019$ & $0.1174 \pm 0.0015$ & $0.1170 \pm 0.0009$ \\\bottomrule
\end{tabular}
\caption{Determinations of the strong coupling at $m_\tau$ for three flavors (middle line) and at $m_Z$ for five flavors (bottom line), based on fits to lattice data at several orders. The reported uncertainty is obtained by adding the individual contributions in quadrature.\label{tab:loops}}
\end{table}

\section{Conclusions}

The perturbative prediction for the static energy is improved by removing the $u=1/2$ renormalon through the MSR mass at a fixed scale $R_0$, and by using R-evolution to resum the associated large logarithms to N$^3$LL accuracy. Ultrasoft logarithms are also resummed to N$^3$LL through the pNRQCD RGE, and we extend the validity of perturbation theory to $r\simeq 0.5\,\mathrm{fm}$ using radius-dependent renormalization-scale profiles. The resulting R-improved static energy includes all $\mathcal{O}(\alpha_s^4)$ terms and both resummations. We conclude that the $\lambda$ parameter that accounts for higher order missing terms in R-evolution must be varied to account for this instability. Profile-function parameters are varied within controlled ranges to ensure $\mu_s>\mu_{\rm us}$, perturbative stability, order-by-order convergence, and a realistic estimate of theory uncertainties. With these ingredients, the full prediction reproduces the expected linear rise of the static energy at large $r$.

For the extraction of $\alpha_s$, we fit our improved pQCD expression to HotQCD data up to \mbox{$r\simeq 0.45\,\mathrm{fm}$}, using an updated value for $r_1$. Energy offsets are treated either as a single parameter or as independent ones for each ensemble; both approaches agree well for $r\geq 0.35\,\mathrm{fm}$, and we restrict our analysis to datasets for which the difference is negligible. Good convergence between one-, two-, and three-loop fits is observed under this restriction. Our final N$^3$LO determination is
\begin{equation}
\alpha_s^{(n_\ell=3)}(m_\tau)=0.3093\pm0.0063\,,
\qquad
\alpha_s^{(n_\ell=5)}(m_Z)=0.1170\pm0.0009\,,
\end{equation}
where the quoted uncertainties incorporate all sources of the error budget. Since only uncorrelated statistical uncertainties are publicly available, the reduced $\chi^2_{\rm min}$ is large (we obtain values around 250), and a shift in the central value or an increase in the total uncertainty may arise once the full covariance matrix becomes available. Nevertheless, our final result is consistent with the PDG world average and previous lattice determinations when using a common value of $r_1$ and a consistent matching prescription.

Future improvements include using the full lattice covariance matrix, subtracting the \mbox{$u=3/2$} renormalon to improve convergence at larger distances, and employing theory nuisance parameters to refine the theoretical covariance matrix.

\begin{acknowledgments}

The author would like to thank P.~Petreczky and J.\,H.~Webber for providing us with HotQCD lattice data and for useful conversations.
JMMV acknowledges funding from European Social Fund Plus, Programa Operativo de Castilla y León and Junta de Castilla y León through Consejería de educación.

\end{acknowledgments}

\bibliographystyle{../../draft/JHEP}
\bibliography{../../draft/NRQCD}

\end{document}